# Observation of Size-dependent Optical Modes and Confinement in Fin Cavities using Angle-Resolved Cathodoluminescence

Kiana Malmir, [1] Jonathan Lee, [2]Douglas Cameron, [2] Babak Nikoobakht, [1] *

*[1] National Institute of Standards and Technology, Gaithersburg, MD 20899, USA.*

*[2] Gatan, Inc., Pleasanton, CA 94588, USA.*

*Babak Nikoobakht, 0000-0003-4369-683X (babak.nikoobakht@nist.gov)*

*Kiana Malmir, 0000-0002-1971-4455*

*Jonathan Lee, 0000-0002-1113-8799*

*Douglas Cameron, 0000-0002-5435-2082*

**Abstract**

We observe novel optical fringes and emission coherence in laterally formed semiconductor fins using wavelength- and angle-resolved cathodoluminescence (WARCL). These findings demonstrate that cavity mode and emission confinement are directly governed by cavity geometry including cavity dimensions, facet angles, and orientation. In flat fins narrowing the width from 3000 nm to 200 nm sharpens fringe intensity, indicating stronger mode confinement and increased coherence. Finite-difference time-domain (FDTD) simulations confirm that the observed fringes originate from internal reflections among the fin facets, consistent with a Fabry-Pérot (FP) cavity model where the cavity length determines the fringe spacing. FDTD also reveals a transition from multi-mode to single-longitudinal-mode behavior in 200-nm fins, consistent with experimentally observed fringe narrowing. Increasing fin length from 10 μm to 30 μm compresses fringe spacing via longer round-trip paths. Conversely, tilting the fin end-facet from 90° to 45° reduces the effective light path length, increasing the incoherent optical emission. Remarkably, this loss can be offset by adjusting the fin length-to-height ratio. Results suggest WARCL is a robust *in-situ* method for direct characterization of individual dielectric nanocavities, mapping their modal profiles and light directionality as a function of geometry for custom design of nanoscale optical sources and waveguides.

Main text

Semiconductor nanostructures with engineered materials and geometries hold great potential for tailoring light-matter interactions, enabling advanced functionalities such as optical field control and modulation at the nanoscale for sensing applications including optical trapping [1–5], advanced photonic devices for micro-light emitting diode (micro-LED) displays, high-speed optical interconnects [6,7], and quantum information technologies [8]. More specifically, semiconductor nanocavities offer tunability across a broad spectral range via compositional and size engineering, suitable for nanoscale light sources [9] and integration into chip-scale devices[10–14]. Beyond these recent advances, the geometric configuration of a nanocavity, namely, its shape, size and orientation on the substrate, remains of paramount importance to the overall cavity performance. For instance, nanowires (NWs) with uniform diameters can support discrete optical modes [13,15], while a gradient in the cavity diameter can suppress higher-order transverse modes, enabling mode switching [16]. This size-dependent property could be tailored for selective excitation of the fundamental optical mode, leading to improved beam directionality within the cavity[17], which in vertically standing NW cavities has been shown to optimize the threshold gain[18].

We utilize WARCL spectroscopy [19,20], a spatially high-resolution microscopy technique for directly measuring angle of the emitted photons[21]. We systematically analyze the emission patterns of laterally formed fins with respect to three primary parameters of fin width, facet angle, and cavity length. To the best of our knowledge, spatial and temporal coherence have been primarily investigated in single metallic nanoparticles and their arrays, for instance, in polariton-based and plasmonic cavities[21–24]. While WARCL has been used for exploring collective properties of photonic bandgaps, e.g., periodic arrays of nanostructures [20,24–26], studies of single dielectric nanocavities remain largely limited to optical excitation and analysis of optical interference patterns, for example, at the end-facets of a nanowire [27]. This is primarily due to a weaker coupling of the high-energy electron beam to the dielectric cavity's electromagnetic modes in CL measurements.

In this work, we investigate the emission coherence of micro- and nanoscale fins of AlGaN and ZnO using a localized electron beam as the excitation source. Lateral fin geometries serve as an ideal model system for next generation on-chip light sources. Compared to free-standing upright nanocavities, they mitigate lattice heating and non-radiative recombination at high injection currents [28–31]. We start with flat fins of AlGaN that are laterally formed on GaN using a selective etching process, previously published[32]. These fins are approximately 500 nm tall and 10 μm long. In this study we choose four different widths of 200 nm, 400 nm, 750 nm, and 3000 nm (Fig. 1, bottom). CL spectroscopy of an AlGaN fin with an Al mole fraction of 0.07 reveals a narrow band-edge emission at approximately 340 nm (see the supplementary material, Fig. S1(a)). While the CL spectrum represents the intensity of the emitted photons and their wavelength (λ) relative to the spatial position of a nanostructure, in a typical WARCL setup (see the supplementary material, Fig. S2), the parabolic mirror in the Scanning Electron Microscope (SEM) chamber simultaneously collects the intensity of the emitted photons as a function of both wavelength (λ) and emission angles (θ and φ). In Fig. 1(top), the angle-resolved CL emission patterns in polar maps are presented for fins (marked as dashed boxes) with different widths. Optical fringes appear as annular rings for all fins. We attribute the appearance of this interference pattern to internal facet reflections and formation of a standing-wave-like pattern at 340 nm. It is also noted that these modes are only seen at high angles, i.e., far from the surface normal direction. The fringe sharpness declines in the wider fins, while the fringe spacing increases with fin width, from ≈ 6° for the 200 nm fin to ≈ 15° for the 3000 nm fin (explained below).

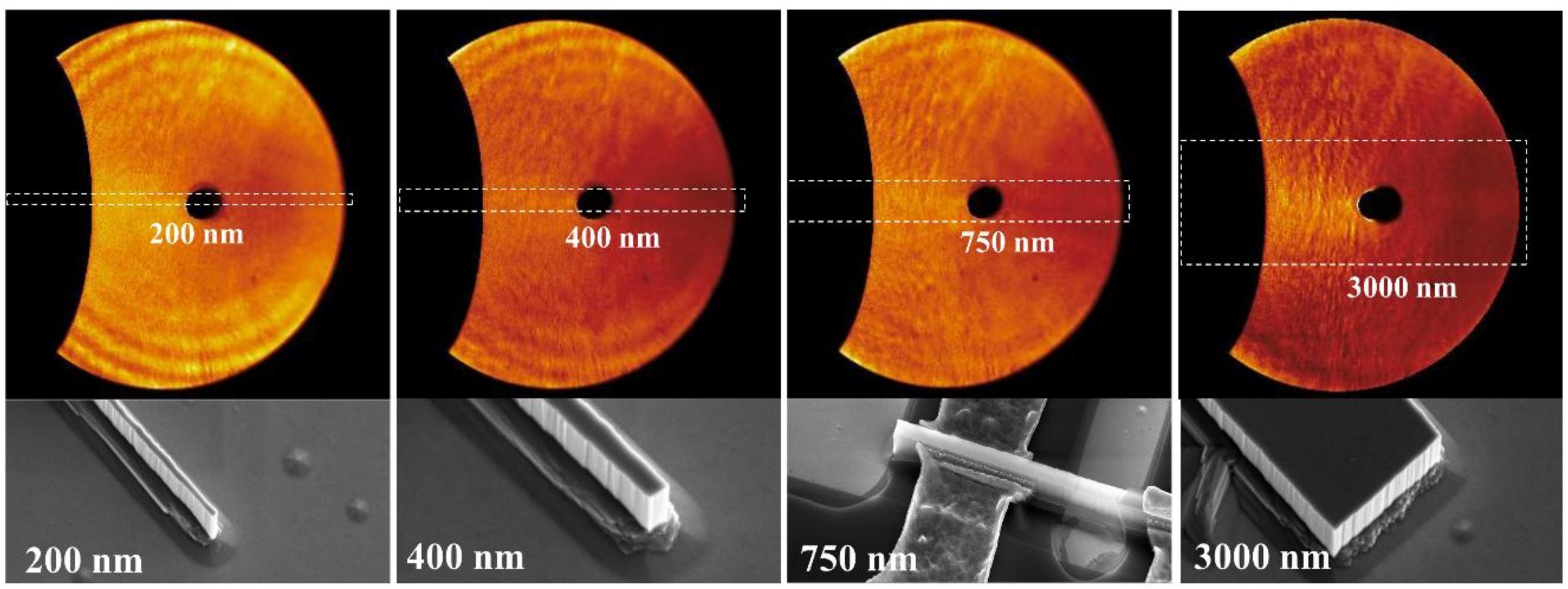

FIG. 1**.** Angle-resolved CL (top) and corresponding SEM images (bottom) of AlGaN fins with four different widths: 200 nm, 400 nm, 750 nm, and 3000 nm (the dashed boxes indicate the fin position and its footprint). All fins are 500 nm tall and 10 µm long. Distinct interference fringes in the ARCL emission patterns are observed for all widths, and the fringe contrast is most pronounced in the narrowest fins.

To further investigate this phenomenon, we conducted finite-difference time-domain (FDTD) simulations of an emitting dipole at λ = 340 nm within an AlGaN fin (see the supplementary material). Figs. 2(a) and 2(b) compare the near-field electric field (E-field) maps of 200 nm and 3000 nm wide fins on GaN substrate. In the 200 nm wide fin, viewed from its end-facet (Fig. 2(a)), a strongly confined E-field within the fin is observed, while in the 3000 nm fin (Fig. 2(b)), the E-field is distributed across the fin width, as transverse resonance modes. Along the fin length, as depicted in Figs. 2(c) and 2(d), these cavity modes are also formed in both narrow and wide fins; however, the mode amplitude is stronger in the narrower one. These modes, consisting of evenly spaced antinodes, suggest interference between forward- and backward-propagating light indicative of partial reflections at the fin end-facets. This is a classic signature of a Fabry-Pérot cavity, where reflections at the end-facets of the dielectric cavity lead to resonance conditions that depend on the wavelength and the effective optical path length.

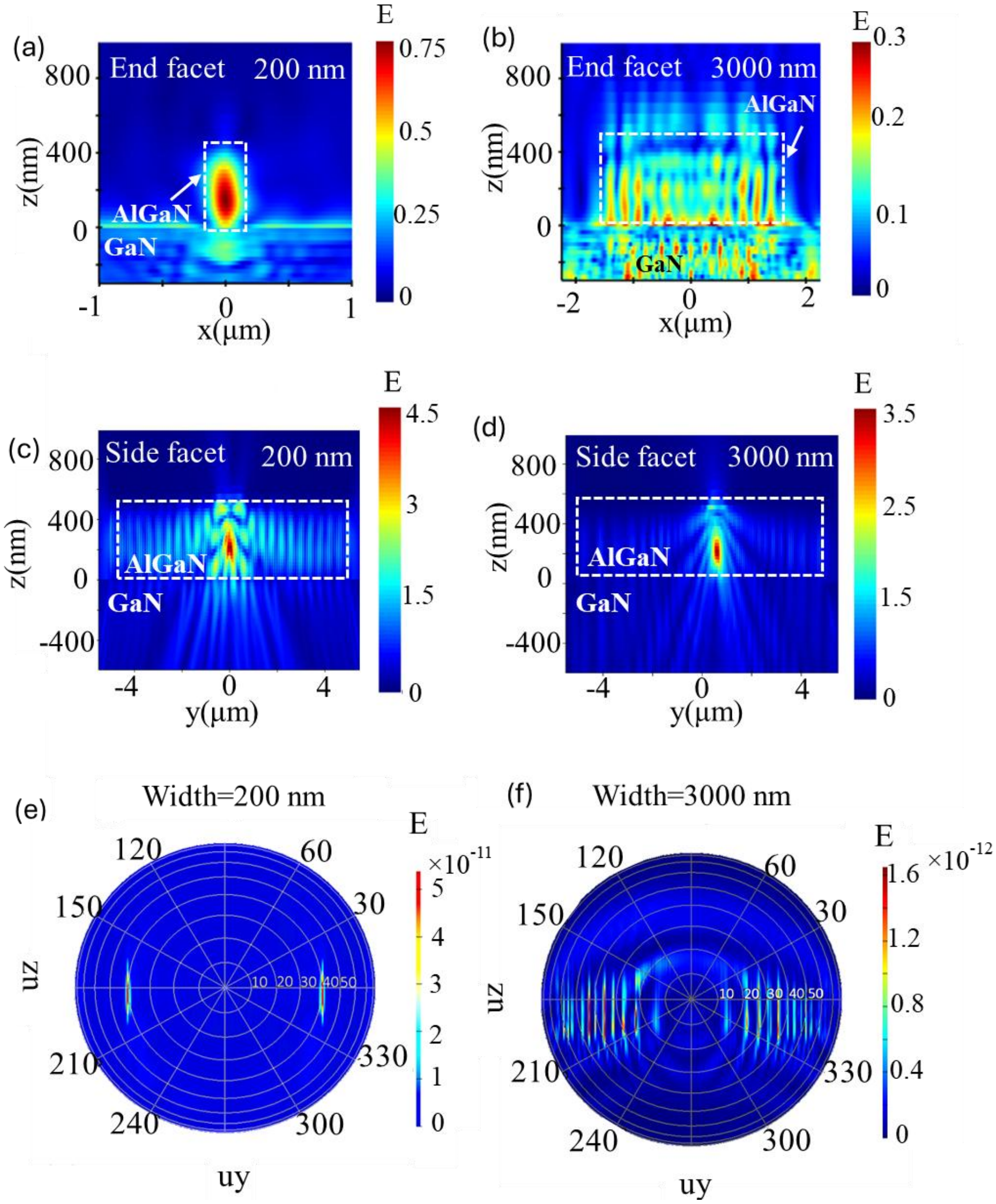

FIG. 2. E-field distributions (arbitrary units) at 340 nm of flat AlGaN fins: 200 nm wide (left column) and 3000 nm wide (right column). The fin (10 μm long, 500 nm tall) is placed on a 2 μm thick GaN slab. The dashed box represents the fin. The emitting dipole with a random orientation is placed 250 nm above the interface. Near-field E-field maps for the narrow and wide fins: (a, b) the end-facet view, and (c, d) the side view along the length. (e, f) show their corresponding far-field polar maps. In the polar maps, rings represent the polar angle ($\theta$) from the surface normal, while the azimuthal angle ($\varphi$) indicates the in-plane angle relative to the fin axis.

In this regime, partial reflections at the fin interfaces create standing optical waves with the resonance condition given by $2nLcos(\theta_m) = m\lambda$, where $n$ is the effective refractive index, L is the cavity length, $\lambda$ is the wavelength, $m$ is an integer mode number, and $\theta_m$ is the internal propagation angle of the $m$-th mode measured from the fin axis. Satisfying this condition leads to constructive interference and the formation of observable angular fringes in the far-field. The far-field maps (Figs. 2(e) and 2(f)) represent the E-field versus the emission angle, where the angle is indicated by the concentric circles, denoted by $u_y$. The fringe spacings are angular far-field spacings obtained from $\mathrm{u_y} = \sin\theta$, where the peak-to-peak spacing in $\mathrm{u_y}$ was converted to angle via $\theta = \arcsin(u_y)$. From the simulations, the fringe spacings are ≈ 5° for the 200 nm wide fin and ≈ 10° for the 3000 nm wide fin, while the corresponding experimental values are ≈ 6° and ≈ 15°, respectively (see the supplementary material, Table SI). In the simple Fabry-Pérot picture, the fringe spacing scales inversely with the effective optical path length, so shorter cavities yield larger fringe separation. However, experimentally, we observe a larger fringe spacing for the 3000 nm wide fin. To address this inconsistency, we propose that the fin length sets the nominal fringe period, while the fin width can alter this spacing. As the fin becomes wider, the optical mode volume increases, and several lateral modes are excited, so the effective coherence length of any single longitudinal mode is reduced by mixing of longitudinal and transverse modes. This effect explains the observation of faint fringes in wider fins and their larger spacing in the ARCL results of Fig. 1. The far-field E-field map of the 200 nm wide fin in Fig. 2(e) shows two prominent narrow lobes (fringes) on each side of the fin. These lobes are attributed to the excitation of resonance modes in the longitudinal direction, indicating a highly directional emission cone in the 40° band (relative to the surface normal). In contrast, the 3000 nm wide fin in Fig. 2(f) shows multiple weak lobes, exhibiting a periodic angular emission. For the 3000 nm wide fin, the increased transverse-mode density reduces the magnitude of the longitudinal resonances, consistent with the weaker modulation and broader angular distribution observed in the simulated far-field line cuts in Figs. S3(a) and S3(b).

To further understand the effective cavity length and coherence of the emission in fins, we explored 160 nm wide fins with tapered top and end facets (bottom of Fig. 3(a)). The tapered ZnO fins were grown on a GaN surface via a previously published surface-directed vapor-liquid-solid growth process[33]. The WARCL map of the ZnO fin in Fig. 3(a) shows distinct interference fringes (white arrows). Similar to the AlGaN fins, these fringes appear at high-angle bands. We attribute this pattern to the internal reflections between the fin facets and the formation of resonance modes, even though the top two facets are tilted. Previously, we showed that the fringe pattern intensifies as the fin height increases [34]. However, the impact of fin tapering on the coherence of the emission remained unclear. The WARCL emission map of the ZnO fin filtered at 380 nm reveals two bright regions located on either side of the fin (dashed box) with an emission angle ($\theta$) that is mostly confined between 15° and 45°.

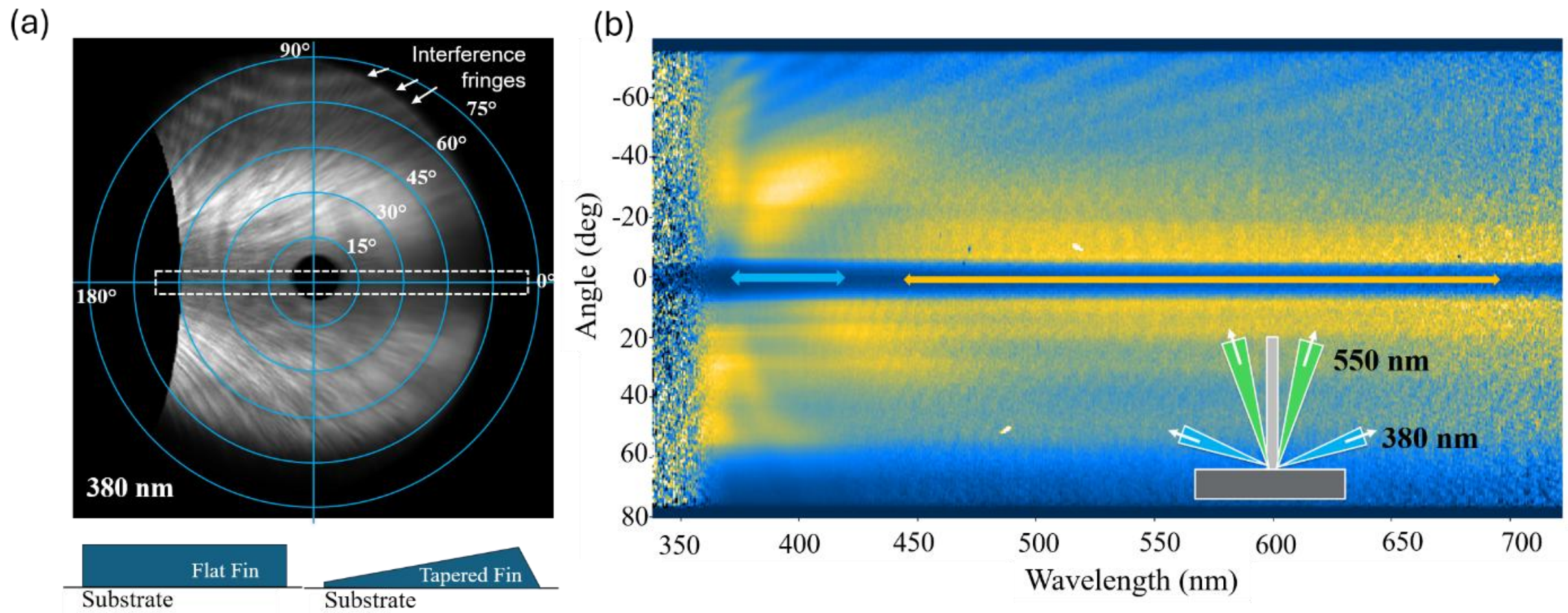

FIG. 3. (a) Wavelength- and angle-resolved CL of a 10 μm long ZnO fin in a polar map filtered at 380 nm. The dashed box indicates the fin orientation. Two bright regions show the emission lobes on the fin sides as well as a series of interference fringes marked with white arrows. Bottom: profiles of flat and tapered fins. (b) WARCL emission intensity of the same fin as a function of wavelength. The light yellow represents the emission at 380 nm with the highest intensity (dark blue is the lowest intensity). This emission band (blue arrow) is more concentrated at higher angles relative to the emission at 550 nm (orange arrow). Inset: Schematic depicts the anisotropy in emission angle shown in the cross section of the fin.

This is part of the fin UV emission that is incoherent and was not detectable in the flat AlGaN fin (Fig. 1). The wavelength distribution of the CL emission in Fig. 3(b) reveals an intense emission band (light yellow color) in the 370 nm to 380 nm range and a broad defect-related emission centered at about 550 nm (marked with blue and orange arrows, respectively). Moreover, the WARCL map shows that the ZnO UV emission (see the supplementary material, Fig. S1(b)) exits the fin at higher angles relative to the defect-related emission. This anisotropy in the emission angle is schematically shown by blue and green cones in the inset of Fig. 3(b).

We next modeled the tapered ZnO fins using FDTD. We chose three fin structures with end-facet angles of 45°, 65°, and 90° and a uniform tapering slope for the top facet, as presented in Fig. 4 (insets). These geometries were selected based on the dominant fin shapes that are obtained via the growth process. The fin base length and height were maintained at 5 μm and 1 μm, respectively. In Figs. 4(a)-4(c), the simulated E-field intensity distributions at the near band-edge emission wavelength (380 nm) are shown for the three geometries. These results show that a tapered ZnO fin exhibits different cavity mode confinement and emission localization as a function of its shape. The standing-wave patterns form along the longitudinal axis of the fins, and the antinodes arise from resonant multiple-beam interference, a signature of Fabry-Pérot cavities.

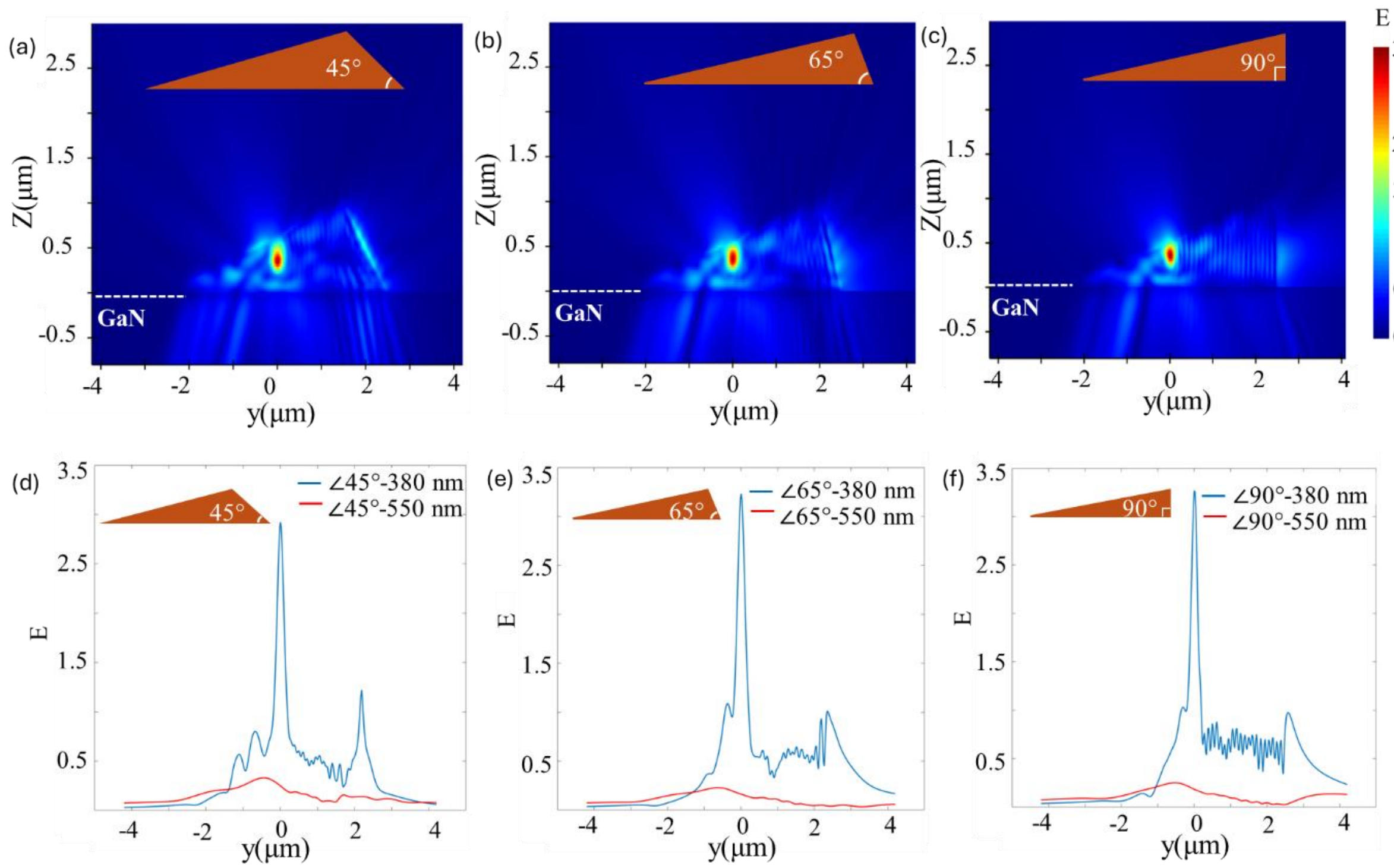

FIG. 4. E-field distributions at 380 nm for tapered ZnO fins with end-facet angles of (a) 45°, (b) 65°, and (c) 90° (inset). The dashed line shows the GaN surface. The hotspot represents the emitting dipole which is randomly oriented and placed 250 nm above the ZnO/GaN interface. (Color bar: normalized E, arbitrary units). (d-f) 1D electric field profiles along the y-axis (at x = -100 nm) for (d) 45°, (e) 65°, and (f) 90° at 380 nm (blue) and 550 nm (red). The refractive index of ZnO is approximately 2 at 380 nm and 1.8 at 550 nm, while that for GaN varies from about 2.5 to 2.3 across the same spectral range.

These results show that the antinodes stay mostly parallel to the end-facet as its angle increases from 45° to 90°. Moreover, the interference patterns within the fin become more intense, indicating stronger reflections at the fin-air and fin-substrate interfaces as shown in the one-dimensional (1D) E-field intensity profiles of Figs. 4(d)-4(f). The FDTD results also show that the field intensity at the near band-edge emission wavelength (blue) is noticeably stronger than that of the defect-related emission wavelength at 550 nm (red), which agrees with the weaker fringes observed (at 550 nm) in the WARCL analysis. In this framework, the fin length is expected to set the basic fringe spacing, while the cavity shape defines the strength of the resonances, light confinement, and coherence, as discussed below.

To further validate the Fabry-Pérot cavity along the fin length, we investigated fringe formation in a 30 μm tapered ZnO fin (1 μm tall; SEM in Figs. 5(a) and 5(b)). Interestingly, its WARCL map (Fig. 5(c)) shows fringes in the high-angle band, consistent with the 380 nm UV photons being emitted at high angles (Fig. 3(b)) and coupling to the fin-cavity modes. The dense fringe spacing follows the expected inverse scaling with cavity length. Moving the excitation box toward the shorter part of the fin (Fig. 5(b)) shifts the fringe positions (data not shown), confirming that the cavity modes track the fin dimensions. FDTD simulation of a similar fin (see the supplementary material, Fig. S4) reproduces the densely spaced fringes (2.9° vs 3.5° measured). All measured and simulated spacings are summarized in Table SI. Figure 5(d) shows that a 30 μm tapered fin (red) emits more strongly at its resonance peak than a comparable flat fin (blue), and its $|E|^2$ time trace shows a shorter oscillatory tail, indicating faster decay and higher optical loss. This is consistent with the larger incoherent emission intensity observed in the tapered ZnO fin (Fig. 3(b)) and its absence in the flat AlGaN fin (Fig. 1). Thus, tapering shortens the effective cavity round-trip path and coherence length.

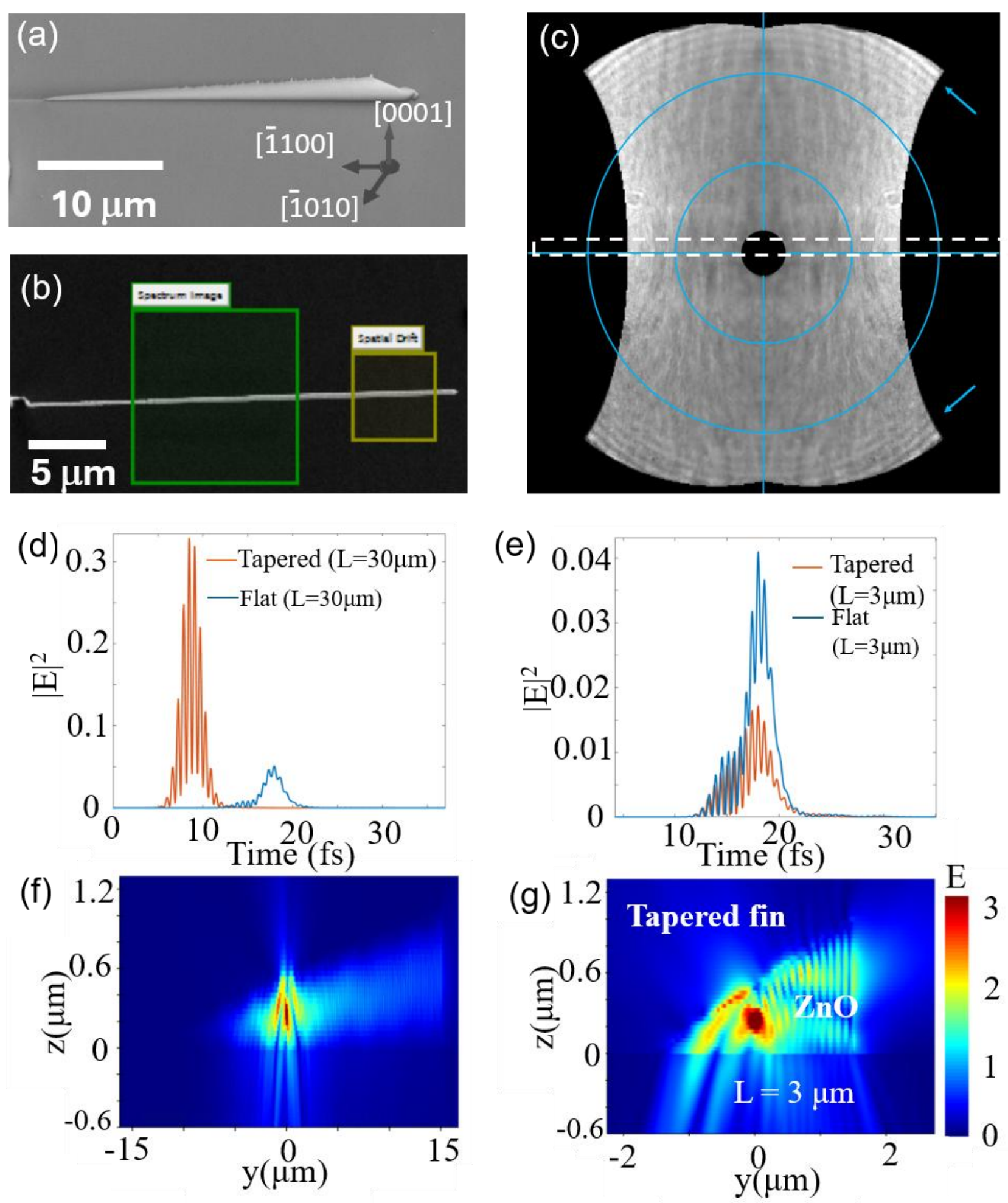

FIG. 5. SEM images of a 30 μm tapered ZnO fin from (a) side- and (b) top-views. (c) ARCL map upon electron-beam excitation; the dashed box indicates the fin position, and the concentric rings denote the emission angle bands with fringes (blue arrows) appearing at high angles relative to the surface normal. (d-g) Effect of tapering on cavity-mode confinement for two different fin lengths: temporal response ($|E|^2$) for (d) L= 30 μm and (e) L= 3 μm and side-view near-field |E| maps at 380 nm for tapered fins with (f) L=30 μm and (g) L=3 μm (shared arbitrary-unit |E| scale).

Below a certain fin length, however, this trend reverses and optical confinement increases. For instance, a 3 μm long tapered fin shows a $|E|^2$ time trace comparable to a flat one (Fig. 5(e)). We believe this is the size regime where the fin length and height are comparable in size, strengthening internal reflections among the end, top, and bottom facets. The near-field |E| maps of Figs. 5(f) and 5(g) confirm stronger amplification in the shorter fin, whereas in a 30 μm fin the larger top facet couples more guided light into the substrate or out of the fin.

Beyond mode confinement and photon lifetime, the end-facet angle also governs the direction and broadening of the emitted beam. The far-field emission (Fig S5) shows that the tapered fin radiates asymmetrically, toward its tall end. To quantify this, the Full Width at Half Maximum (FWHM) of the main emission lobe (central interference fringe) was extracted from the far-field radiation patterns (see the supplementary material). Making the end-facet more vertical reduces the FWHM from 0.09 at 45° to 0.06 at 90° (see the supplementary material, Table SII) indicating a narrower and more directional beam. This sharpness can be attributed to a higher quality factor $Q = \frac{\lambda}{\Delta\lambda}$, where ($\lambda$ is the resonance wavelength and $\Delta\lambda$ is the resonance linewidth, consistent with stronger Fabry-Pérot-like feedback in the flat fin. This behavior arises because a vertical (90°) end-facet reflects the partially guided light back into the fin, reinforcing a stronger standing-wave pattern and improving modal confinement, whereas a higher tilt angle facet directs much of the light downward into the substrate weakening the feedback and broadening the far-field emission.

In conclusion, the impact of fin geometry on the cavity-mode formation, optical confinement and emission characteristics was investigated in individual semiconductor micro- and nanofins using wavelength- and angle-resolved CL spectroscopy. Results revealed size- and shape-dependent optical fringes, which are attributed to internal reflections between the fin facets and the presence of a Fabry-Pérot behavior. WARCL measurement of flat fins showed that in narrower fins, the fringe patterns became increasingly stronger, indicating stronger optical mode confinement in agreement with near- and far-field FDTD simulations. Wider fins exhibited larger optical mode volumes, resulting in increased coupling between longitudinal and transverse modes and a reduced effective coherence length. WARCL spectroscopy further showed that both flat and tapered fins emit near-band-edge radiation predominantly into a high-angle band, in agreement with the far-field FDTD calculations. However, tapered fins showed a more pronounced incoherent near-band-edge emission. These results suggest that the end-facet angle and cavity width can significantly affect the light round-trip path, photon confinement and escape. These findings demonstrate the critical role of nanocavity geometry in controlling optical modes and provide a pathway for optimizing optical gain, cavity quality factor, light extraction efficiency and beam directionality in integrated photonic applications.

## Supplementary Material

See the supplementary material for additional CL spectra, near- and far-field line cuts, far-field radiation patterns, the angular fringe-spacing table, and the main-lobe FWHM table (Figs. S1–S4 and Tables SI and SII).

## Acknowledgments

This work was partially performed with funding from the CHIPS Metrology Program, part of CHIPS for America, National Institute of Standards and Technology, U.S. Department of Commerce.

## Author Declarations

Conflict of Interest**:** The authors declare no conflict of interest.

K. M.: Investigation, Formal analysis, Methodology, Visualization, Writing – Review & Editing. KM performed the simulations, revised the figures, and edited the manuscript.

J. L. and D. C.: Investigation. Equally performed the spectroscopy measurements and reviewed the manuscript.

B.N.: Conceptualization, Formal analysis, Visualization, Supervision, Writing – Original Draft. BN conceived the project, analyzed the experimental data, prepared the figures, supervised the research, and wrote the original draft of the manuscript.

## Data Availability Statement

All data in support of the findings of this paper are available within this article and the supplementary material.

# SUPPLEMENTARY MATERIAL

# Observation of Size-dependent Optical Modes and Confinement in Fin Cavities using Angle-Resolved Cathodoluminescence

Kiana Malmir, [1] Jonathan Lee, [2]Douglas Cameron, [2] Babak Nikoobakht, [1] *

[1] *National Institute of Standards and Technology, Gaithersburg, MD 20899, USA.*

[2] *Gatan, Inc., Pleasanton, CA 94588, USA.*

*Babak Nikoobakht, 0000-0003-4369-683X (babak.nikoobakht@nist.gov)*

*Kiana Malmir, 0000-0002-1971-4455*

*Jonathan Lee, 0000-0002-1113-8799*

*Douglas Cameron, 0000-0002-5435-2082*

Fig. S1(a) shows the CL emission spectra of 200 nm and 3000 nm wide AlGaN fins. The 200 nm fin (black curve) has a much stronger near-band-edge peak than the 3000 nm fin (orange curve), and its broad higher-wavelength background is also more intense. This means that the narrow fin couples the electron-beam excitation more efficiently into the guided optical mode and radiates it out, while the wide fin loses more energy into weak, leaky modes. The overall trend matches our simulations and analysis in the main text. Namely, narrow fins give better optical confinement and stronger emission, whereas wider fins show reduced confinement and much weaker output. Fig. S1(b) presents the CL spectrum of a ZnO fin at room temperature (295 K), showing an intense near-band-edge UV peak at ~380 nm and a broad, weaker defect-related emission band centered near 550 nm, consistent with the emission bands identified in the main text (Fig. 3(b)).

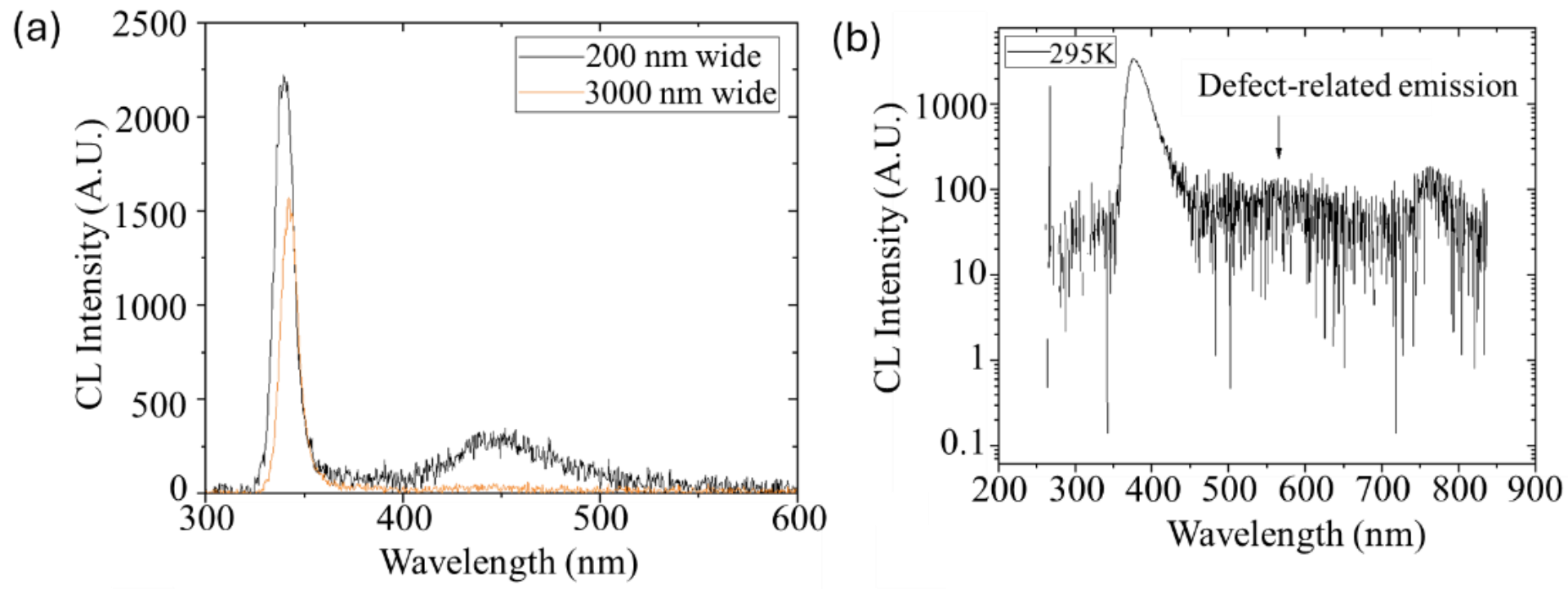

FIG. S1**.** CL emission spectra for (a) 200 nm and 3000 nm wide AlGaN fins and (b) ZnO fin at room temperature.

## WARCL spectroscopy

Conventional CL spectroscopy measures wavelength-resolved CL of the emitted light (Intensity vs. wavelength) over the specimen. However, WARCL also resolves the CL by emission angle, i.e., the direction in which light is emitted (Intensity vs. angle), in addition to wavelength. For these measurements, the CL emission was analyzed using a Gatan Monarc Pro* equipped with wavelength- and angle-resolved CL capability. The collection geometry is shown schematically in Fig. S2. The resulting dataset, $I(\lambda, \theta, \varphi)$, records the emitted intensity as a function of wavelength and emission direction, where $I$ is intensity, $\theta$ is the elevation (relative to the z direction) and $\varphi$ is the azimuthal direction (in the *x-y* plane). This angular-spectral mapping allows us to understand the novel optical properties of nanoscale emitters, such as the existence of optical modes, their angle of emission, and their dependence on the cavity size and geometry.

* Certain commercial equipment, instruments, materials, or software are identified in this paper to foster understanding. Such identification does not imply recommendation or endorsement by the National Institute of Standards and Technology, nor does it imply that the materials or equipment identified are necessarily the best available for the purpose.

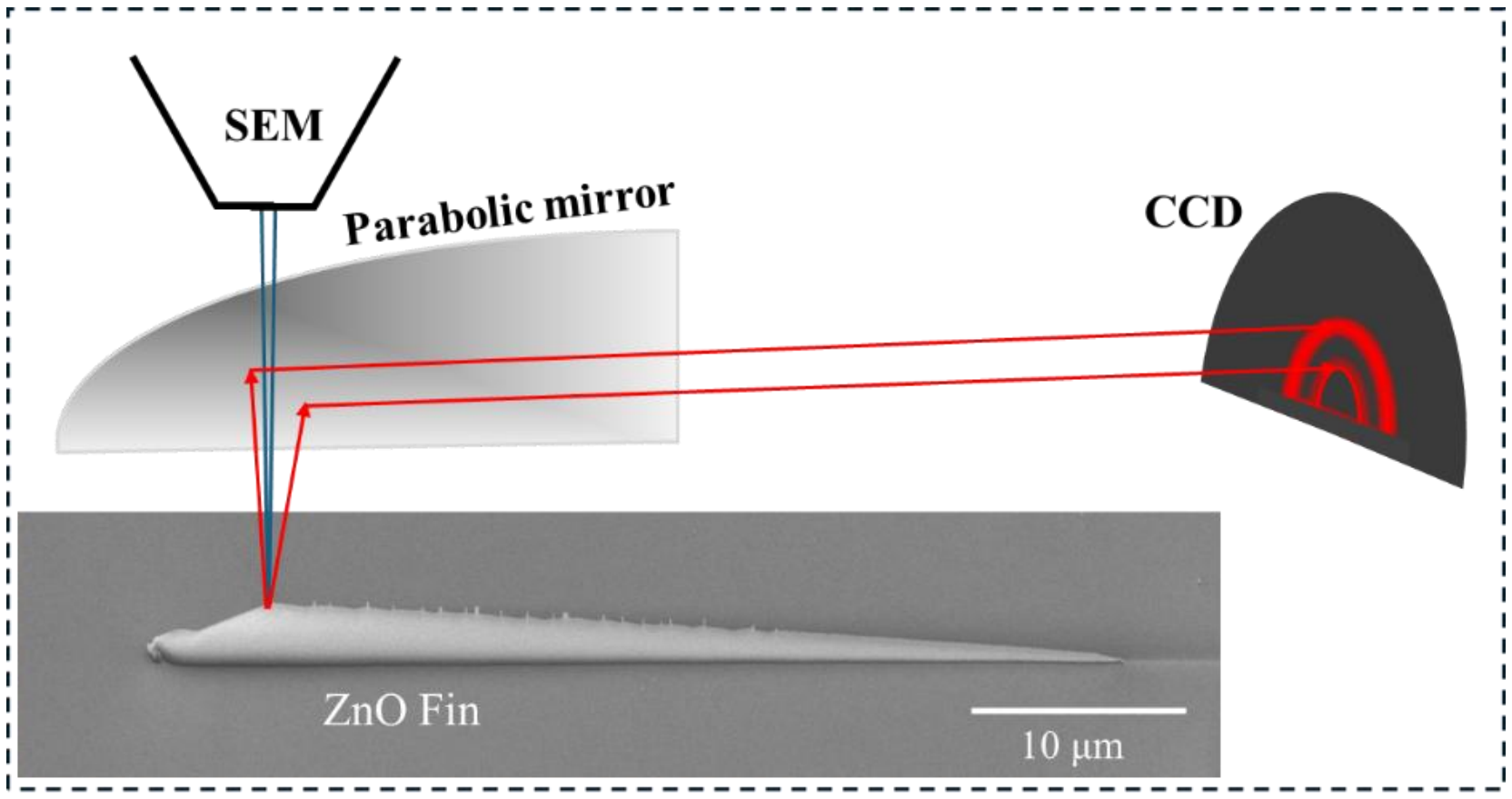

FIG. S2. Schematic of the CL collection setup upon excitation of a tapered ZnO fin by the electron beam. The light collected by the parabolic mirror is received by a Charge-Coupled Device (CCD) for wavelength- or angle-resolved spectroscopy.

Finite-Difference Time-Domain (FDTD) electromagnetic simulations were used to model the full-wave evolution of the electric field in both time and space, capturing both propagating and evanescent components to obtain the energy distribution maps. Simulations of the electric field in steady-state were carried out using ANSYS Lumerical* FDTD with a radiating electric dipole source within the fin. The dipole was randomly oriented at or above the fin-substrate interface and represents the electron-hole (e-h) recombination centers generated by the electron-beam excitation. In these calculations, ZnO, AlGaN, and GaN were modeled as wavelength-dependent dispersive materials using literature refractive index data implemented via Lorentz-Drude fits. A 2 % to 3 % uncertainty in the refractive index is expected due to growth-condition variations but only results in minor spectral shifts. Perfectly matched layers (PML) and refined meshing were used to ensure accurate mode confinement and far-field extraction.

Table SI summarizes the experimental (ARCL) and simulated (FDTD) angular fringe spacings for the fin geometries discussed in the main text. For the flat AlGaN fins, the fringe spacing increases with fin width, from ≈6.4° at 200 nm to ≈15° at 3000 nm, whereas the tapered ZnO fin (160 nm) exhibits the smallest spacing (≈3.5°). The simulated spacings follow the same trend and are systematically smaller than the measured values (by roughly 15–30%), which we attribute to the idealized point-dipole source and the literature dispersion data used in the model. The agreement between the measured and simulated spacings corroborates the Fabry-Pérot origin of the fringes described in the main text.

TABLE SI. Experimental and simulated angular fringe spacings (in degrees), extracted from ARCL maps and from FDTD top-view far-field line cuts along $u_y$.

| Fin Shape | Fin Material | Fin Width (nm) | Fringe Spacing, Experiment (°) | Fringe Spacing, Simulation*(°) |
|---|---|---|---|---|
| Flat | AlGaN | 3000 | 15 | 10.3 |
| Flat | AlGaN | 200 | 6.4 | 4.6 |
| Tapered | ZnO | 160 | 3.5 | 2.9 |
| *In the simulations, fringe spacings are extracted from the top-view far-field maps $E(u_x, u_y)$ by taking a line cut along $u_y$ through the main emission band and converting the peak-to-peak spacing to angle via $\theta = \arcsin(u_y)$. | | | | |

To further validate the role of cavity geometry in shaping mode confinement and emission behavior, we provide additional near- and far-field results. Fig. S3 shows line cuts of the electric field extracted from the end-facet-view maps in Fig. 2 and highlights that the narrow fin sustains a strongly localized and intense mode along its length, consistent with the trends discussed in the main text. The near-field line cuts in Fig. S3(a), taken along y at a fixed height (200 nm), confirm that the mode peaks at a similar position along the fin for both structures, with a higher peak amplitude for the narrow fin. The corresponding side-view far-field patterns (Fig. S3(b)) show pronounced, narrow lobes for the 200 nm fin, whereas emission from the 3000 nm fin is comparatively weak and suppressed across $u_y$. This indicates that the stronger lateral confinement in the narrower fin enhances coupling into well-defined radiation channels and yields more directional out-coupling.

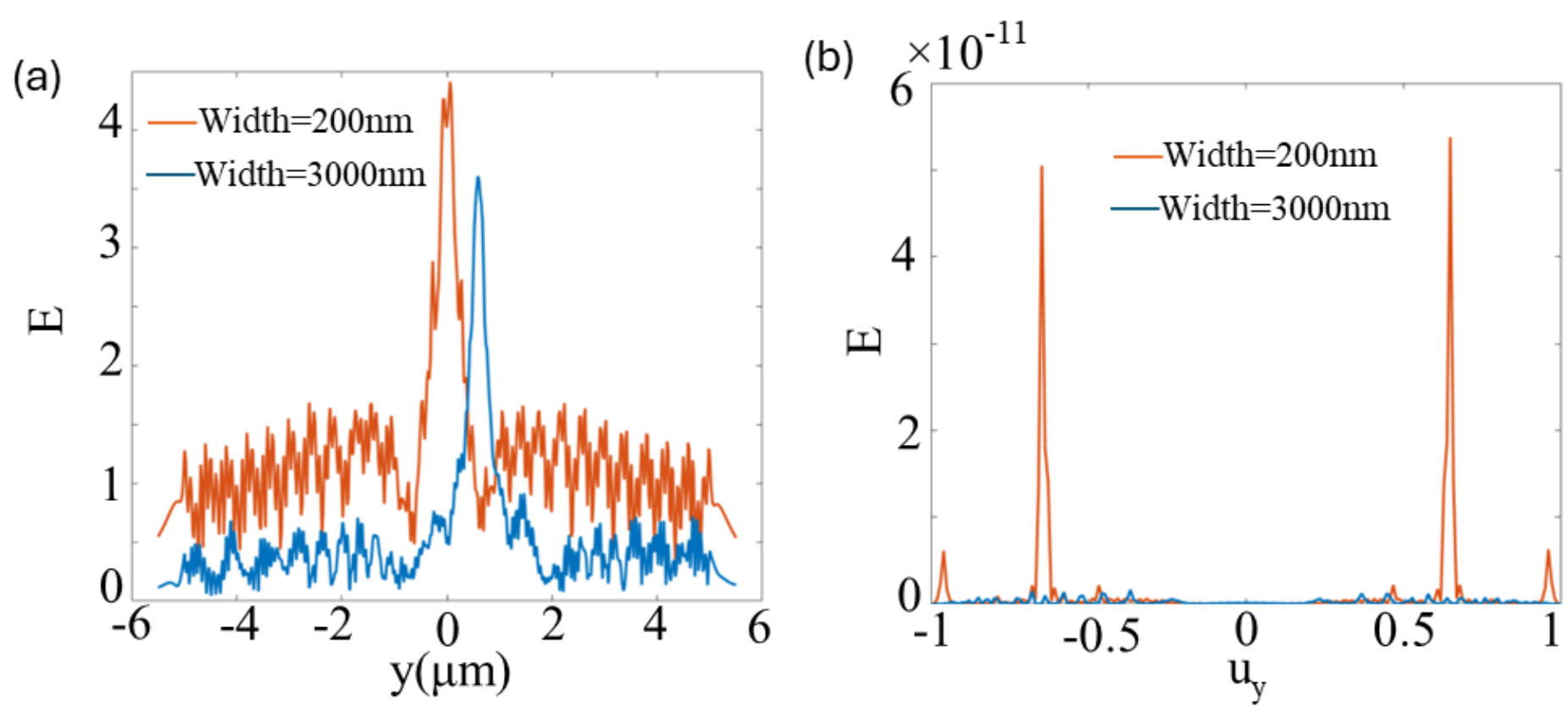

FIG. S3. Near- and far-field comparison for non-tapered (flat) fins with 200 nm (red) and 3000 nm (blue) widths. (a) 1D line cut of the near-field E along the cavity length y. (b) 1D line cut of the far-field along $u_y$. Line cuts are extracted from the maps of Fig. 2.

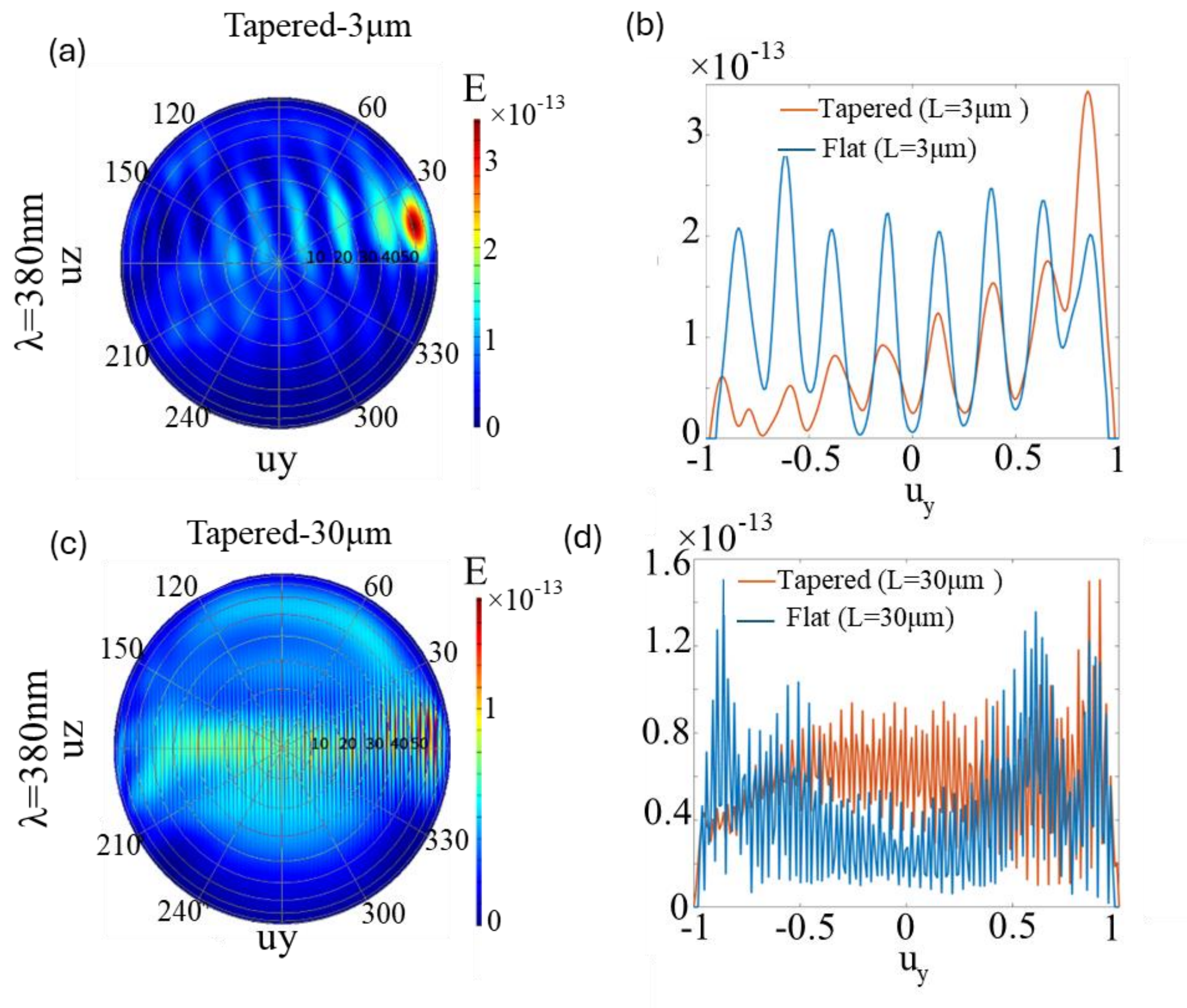

FIG. S4. FDTD E-field (arbitrary units) at far-field of tapered ZnO fins for two cavity lengths. (a) Polar map for tapered fin with L=3 μm. (b) 1D far-field line cut along $u_y$ for the tapered fin (red) and its comparison with a flat fin (blue). (c) Polar map of a 30 μm tapered fin. (d) 1D far-field line cut along $u_y$ for the tapered (red) and its comparison with a flat fin (blue).

Figs. S4(a) and S4(b) compare the side-view far-field E of 3 μm tapered and flat fins. The tapered fin (Fig. S4(a)) exhibits a directional pattern, with radiation concentrated into a dominant lobe at positive $u_y$. The 1D cut in Fig. S4(b) confirms that tapering concentrates the radiated field into a single dominant, off-axis lobe at positive $u_y$, whereas the flat fin produces pronounced oscillatory emission over a broad range of $u_y$. Figs. S4(c) and S4(d) show the same comparison for 30 μm fins. The longer fin produces a much denser interference pattern in both cases, as expected for a larger effective aperture. The tapered fin (Fig. S4(c)) shows a dense but apodized fringe pattern with smoother envelope, while the 1D cut in Fig. S4(d) confirms strongly oscillatory emission for both designs. The flat fin exhibits stronger, higher-contrast fringes, whereas the tapered case shows fewer prominent peaks across much of $u_y$. It is noted that the fringe (resonance) sharpness is greater in the flat 30 μm fin than in the tapered one.

Table SII summarizes the FWHM of the main far-field emission lobe (in $u_y$) for end-facet tilt angles of 45°, 65°, and 90° at 380 nm and 550 nm.

TABLE SII. FWHM of main emission lobe for different fin facet tilt angles, expressed as the direction cosine $u_y = sin\theta$.

| Wavelength | 45° | 65° | 90° |
|---|---|---|---|
| 380 nm | 0.09 | 0.07 | 0.06 |
| 550 nm | 0.11 | 0.09 | 0.08 |

Regarding the emission direction and beaming of light to free space, the far-field emission characteristics further clarify another role of the fin shape. The far-field polar map of Fig. S5(a) shows that the tapered fin exhibits a more directional pattern, producing pronounced oscillatory emission over a broad range of $u_y$. In the case of a fin with an upright end-facet (90°), the emission is asymmetric and well directed toward the tall end of the fin and along the fin long axis. Irrespective of the fin length, tapered fins radiate into a similar high-angle band (Fig. S4) in agreement with the WARCL data. At 65° tilt (Fig. S5(b)), the azimuthal angle of the emission direction shows about 30° rotation relative to the fin long axis. At a 45° end-tilt (Fig. S5(c)), the fin shows a limited optical confinement, indicating a diffuse and weak directionality at both 380 nm and 550 nm.

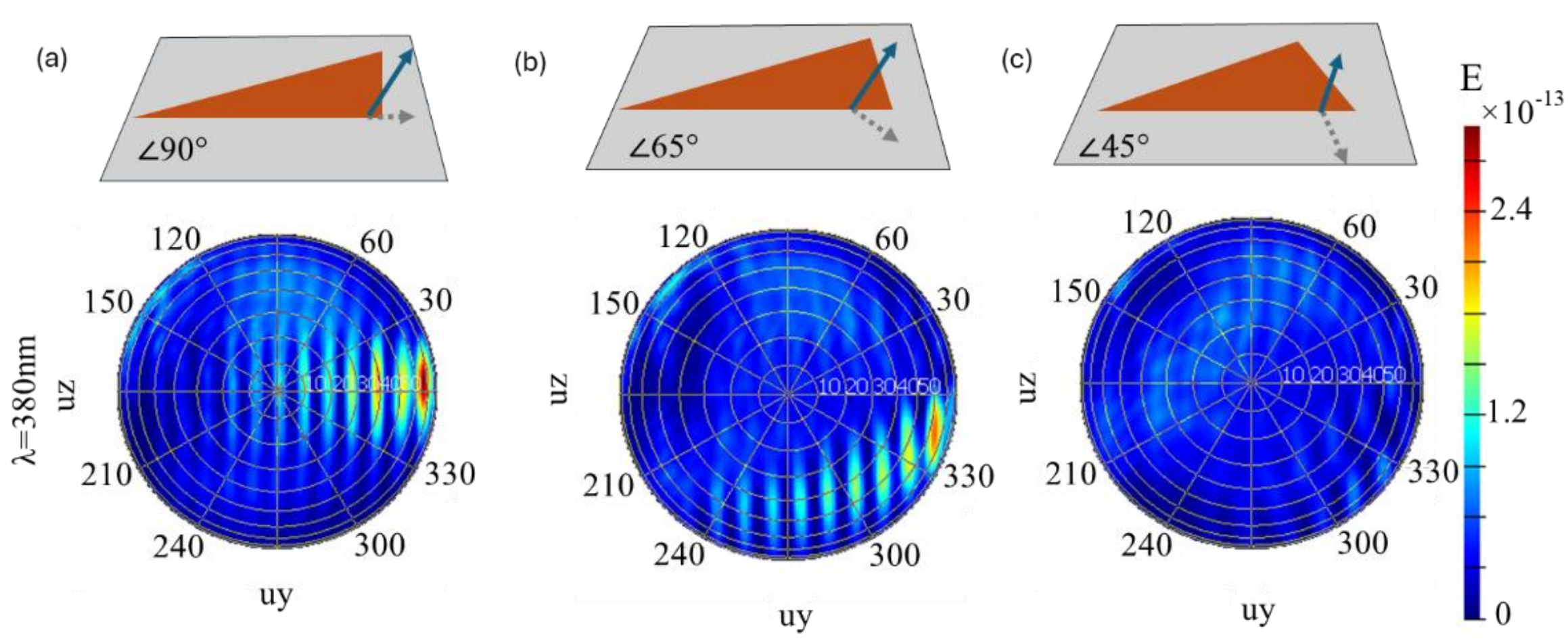

FIG. S5. Simulated far-field polar emission maps at 380 nm for (a) 90°, (b) 65°, and (c) 45° end-facet angles. Asymmetric lobes are observed for all facet angles, indicating that the fin radiates from its tall end. In the insets, the solid blue arrow indicates the dominant emission direction and the gray dashed arrow its in-plane projection; as the end-facet angle decreases, the emission steers away from the fin's long axis. The color scale shows the normalized electric field (arbitrary units) for (a), (b), and (c).